\documentclass[showpacs,aps]{revtex4}
\usepackage{graphicx,verbatim}
\usepackage{epstopdf}
\usepackage{slashed}
\usepackage{comment}
\usepackage{appendix}
\begin{document}

\title{Exploring dynamical gluon mass generation in three dimensions}
\author{John M. Cornwall\footnote{Email cornwall@physics.ucla.edu}}
 \affiliation{Department  of Physics and Astronomy, University of California,
Los Angeles CA 90095}

\date{\today}
 
\begin{abstract}
We re-examine the d=3 dynamical gluon mass problem in pure-glue non-Abelian $SU(N)$ gauge theories, paying particular attention to the observed (in Landau gauge) violation of positivity for the spectral function of the gluon propagator. This is expressed as a large bulge in the propagator at small momentum, due to the d=3 avatar of asymptotic freedom.  Mass is defined through $m^{-2}=\Delta (p=0)$, where $\Delta(p)$ is the scalar function for the gluon propagator in some chosen gauge; it is not a pole mass and is generally gauge-dependent, except in the gauge-invariant Pinch Technique (PT).  We  truncate the PT equations with a recently-proposed method called the vertex paradigm that automatically satisfies the QED-like Ward identity relating the 3-gluon PT vertex function with the PT propagator.  The mass is determined by a homogeneous Bethe-Salpeter equation involving this vertex and propagator.  This gap equation also encapsulates the Bethe-Salpeter equation for the massless scalar excitations, essentially Nambu-Goldstone fields, that necessarily accompany gauge-invariant gluon mass.  The problem is to find a good approximate value for $m$ and at the same time explain the bulge, which by itself leads, in the gap equation for the gluon mass, to excessively large values for the mass. Our point is not to give a high-accuracy determination of $m$ but to clarify the way in which the propagator bulge and a fairly accurate estimate of $m$ can co-exist, and we use various approximations that illustrate the underlying mechanisms.  The most critical point is to satisfy the Ward identity.  In the PT we   estimate a gauge-invariant dynamical gluon mass of $m \approx Ng^2/(2.48 \pi)$.   We translate these results to the Landau gauge using a background-quantum identity involving a dynamical quantity $\kappa$ such that $m=\kappa m_L$, where $m_L^{-2}\equiv \Delta_L(p=0)$.  Given our estimates for $m,\kappa$ the relation  is fortuitously well-satisfied for $SU(2)$ lattice data.
\end{abstract}

 \maketitle

\section{Introduction: The d=3 and d=4 gluon mass problems}

This paper considers dynamical gluon mass generation in a pure-glue d=3 non-Abelian gauge theory (NAGT), based on the Pinch Technique (PT).   Recall that the PT algorithm was introduced \cite{corna9,corn76,corn85,corn114,corn90,cornbinpap} to generate gauge-invariant Green's functions in non-Abelian gauge theories such as an NAGT, and was later extended in d=4 to an algorithm for Green's functions both gauge-invariant and renormalization-group invariant (RGI) \cite{corn141,corn144,corn147}.  Of course, the gluon mass is not a pole mass, or we would see gluons in experiments; it is more in the nature of a screening mass, analogous to the polaron of condensed matter physics---an electron or hole made heavy by coupling to the ionic background.  In an NAGT the gluon couples to a background of other virtual gluons. We give a precise definition to the mass concept, defining a zero-momentum mass $m$ by
\begin{equation}
\label{massdef}
m^2\equiv \widehat{\Delta}^{-1}(p=0).
\end{equation}
(Throughout this paper, hatted quantities are PT quantities.)
We similarly define a Landau-gauge mass by
\begin{equation}
\label{landmass}
m_L^2 \equiv \Delta_L ^{-1}(p=0).
\end{equation}
Here $\widehat{\Delta}(p)$ is the scalar function for the PT gluon propagator and similarly for Landau gauge.   

The PT mass is the zero-momentum value of a running mass $m(p)$ that vanishes like $1/p^2$ at large momentum; see Sec.~\ref{mix}. For technical reasons, in d=3 it is much simpler to ignore the running of the mass, which we do throughout this paper.  Clearly this definition of mass makes sense only if the right-hand side of Eqns.~(\ref{massdef},\ref{landmass}) is finite and positive.  As we discuss at the end of this section  all lattice simulations in the Landau gauge show that this is indeed true.    In Sec.~\ref{landau} we invoke a background-quantum identity showing that the zero-momentum Landau-gauge propagator $\Delta_L(p=0)$ is a finite and positive multiple $\kappa_L^2$ of  $\widehat{\Delta}(p=0)$:
\begin{equation}
\label{bqident}
\Delta_L(p=0)=\kappa_L ^2\widehat{\Delta}(p=0),
\end{equation}
and so $m=\kappa_L m_L$.  Since $\kappa_L < 1$, these two masses are different.  This is to be expected, since the Landau-gauge propagator is gauge-dependent and unphysical; the PT mass as defined in Eq.~(\ref{massdef}) is gauge-invariant.  By estimating $\kappa_L$ and  using the estimate of $m^2$ from the present work we find a fortuitously close agreement between our resulting approximate value of $m^2$   and the Landau-gauge mass  inferred from  simulations.  Or conversely we may take the simulation value $m_L$ and infer $m$, again with fortuitously good agreement, considerably better than the 20-25\% error that we are probably making in our approximate formulation.

  One might think that d=3 mass generation should be an easier problem than in d=4, where renormalization is required.  In contrast,  a d=3 NAGT is superrenormalizable, so that at infinite momentum the gluon coupling $g^2$  does  not change from its value in the classical action.  Nevertheless, we can define a running coupling $\bar{g}^2(p)$ without reference to a renormalization group and this running has  important consequences for the gluon mass in d=3.

A number of theoretical works on the gluon mass in d=3  date from the nineties \cite{corna9,corn76,corn85,corn114,corn90,alexnair,corn120,buchphil,buchphil2,jackpi,jackpi2,nair}. These gave reasonable results for the gluon mass, but a closer analysis \cite{corn120} of the theoretical gluon propagators  seemed to be disappointing and unphysical  for a reason that was not really appreciated at the time:  Non-positivity of the propagator spectral function. This is manifested by a bulge in the Euclidean propagator, clearly evident in Landau-gauge lattice simulations.  Non-positivity  is a consequence of d=3 infrared slavery, inherited from the ``wrong" sign of d=4 asymptotic freedom (AF).  

Aside from the works referenced above, there is also a decade-later work  \cite{agbinpap} using a special form of the PT and oriented to later Landau-gauge lattice data.  The general approach is similar to what is used here, including the addition of massless scalars to the 3-gluon vertex  \cite{corn90}. The massless scalar fields are essentially Nambu-Goldstone (NG) excitations, and must exist as bound states if the gluon mass is to be gauge-invariant with no elementary Higgs fields.   
  Using free vertices and free massive input propagators, the authors   find important non-positivity in the output propagator.   However, there are significant differences from the present work in the treatment of mixing the massless scalars with gluons and of determination of the gluon mass.  Moreover, there is no discussion of the effects of non-positivity on the 3-gluon vertex, which we estimate to be considerable and in the direction of cancelling non-positivity effects in the propagator when used in the gluon-mass gap equation.  Ref. \cite{agbinpap} and other later works benefited from good lattice data on the Landau-gauge propagator\cite{ctt,cucchmen,maas} that we  will recap in Sec.~\ref{crit}.  
The lattice evidence for dynamical generation of some sort of gluon mass is unequivocal in d=3:  The Landau-gauge inverse propagator is not zero at zero momentum, but finite and positive (see Fig.~\ref{maasfig} in the next section).
In d=4 there is also an abundance of lattice work that confirms gluon mass generation. See, for example, \cite{alex1,alex2,bmmp,bouc1,bowm1,bouc2,silva1,silva2,bogo1,bogo3,stern,bogo2,born,
olivsilv,oliv,silvoliv,bowm2,ilge2,cucch1,cucchmend,cucch2,cucch3,bouc3,aouane,gongyo,
silvb,pawsp,zhang}.  Moreover, much is being done in continuum studies of the d=4 problem, mostly by Papavassiliou and collaborators (for a discussion of work up to 2011, see \cite{cornbinpap}; later work can be traced from, for example, \cite{pap14}).  

 In the present paper we argue that in the homogeneous Bethe-Salpeter (B-S) equation governing the value of the dynamical mass, this positivity problem is largely ameliorated by a compensating dip in the 3-gluon vertex, so that the predicted gluon mass value is much less affected by non-positivity than the propagator itself is.  That there must be a dip in the vertex that (partially) compensates the propagator bulge follows from the QED-like Ward identity (see Sec.~\ref{vp1}) relating them.     Approximations not accounting for both the propagator bulge and the vertex dip can give gluons mass values far removed from reality.  For this reason, it is particularly important that the Ward identity be satisfied, even in the face of approximations.  The vertex paradigm that we use here is based on constructing an approximate vertex from which the propagator is extracted using the Ward identity.  Much of the present work is devoted to the study of this complicated non-linear problem.  We can find an explicit Feynman-parameter integral for the approximate vertex at one dressed loop, and it is easy to find the corresponding propagator from the Ward identity.  

Evaluating the many terms of this Feynman-parameter integral to find the vertex itself is a daunting task, never done even for the one-loop perturbative vertex. Even if it were evaluated precisely, it is an approximation that is not necessarily highly accurate. Consequently, we resort to other approximations based on d=5,6 scalar field theories to capture the essence of how the Ward identity relates a propagator bulge to a vertex dip \cite{corn141,corn144,corn147}.  These   are useful because $\phi^3_6$ is asymptotically-free and its d=5 descendant behaves much like a d=3 NAGT; they are reviewed in Appendix \ref{vpapp}.  The models are tweaked so that their one-dressed-loop Schwinger-Dyson equations (SDE) resemble those of an NAGT as much as possible, and to this end some of the fields are endowed with an Abelian charge and corresponding vertex with its Ward identity.  It is uncomplicated to carry out the vertex paradigm construction for these scalar field theories, and the results for the propagator are surprisingly close to previous approximations to NAGTs \cite{corn147}.

\subsection{The vertex paradigm}

The vertex paradigm was   previously used in d=4 for truncating the PT Schwinger-Dyson equations   \cite{corn141,corn144,corn147}.   It begins with  analytic tree-level approximations to the full PT inverse propagator and 3-gluon PT vertex (the inputs) that are massive and therefore free of IR singularities. If the masses do not run, these inputs       exactly satisfy the ghost-free QED-like Ward identity relating  them.   This is a critical point in showing that a one-dressed-loop output approximation to the 3-gluon vertex using these input Green's functions  satisfies the Ward identity.   We then simply apply the Ward identity   to find   the output gluon propagator.  There are a number of technical obstacles to overcome, in particular the treatment of bound-state massless scalar excitations, akin to NG fields, that are necessary if the gluon has mass.  We give a road map of the vertex paradigm in the  Appendices, and more details are found in \cite{corn147}.  In principle, the output Green's functions can be recycled and  used as input functions for another round, but nothing is known about what happens in this second stage.

In reality, the output gluon mass is a running mass, depending on momentum \cite{corn76,lavelle} in both d=3 and d=4.  The author does not know how to guarantee the Ward identity with a momentum-dependent input mass, but it is much easier, although not trivial, if the mass does not run.  In d=4 using a constant mass prevents us from actually finding a value for the mass, which must vanish at infinite momentum for the gluon-mass gap equation to be UV-finite.  (If a  truly constant   bare mass led to a UV-finite solution of the gap equation, then an NAGT with a mass term would be renormalizable in d=4.)  But in d=3 we can still solve the gap equation with a constant mass.  The error made in this approximation is small, since the UV region contributes little to the gap equation.  Ultimately, this   is inconsistent,  because in d=3 a constant mass input to the gap equation automatically leads to an output mass that runs to zero as $1/p^2$ at large momentum.  An identically-vanishing mass is not a solution to the gap equation, which becomes IR-singular in this limit.

Unless otherwise specified, we carry out the vertex paradigm in Euclidean space with the usual Euclidean metric.  We reserve a study of the properties of the dynamical mass in Minkowski space, where it is surely not a pole mass, for the future.   

\subsection{Organization of the paper}

Section \ref{crit} brings up the critical properties, related to infrared confinement, that are the central themes of this paper.
In Section  \ref{vp1} we introduce the vertex paradigm and the Ward identity, and discuss the massless scalar poles necessary for gauge-invariant gluon mass generation.    Section \ref{step1} is a straightforward transcription of earlier d=4 efforts in the vertex paradigm to construct the d=3 pole-free vertex.  Section \ref{mix} constructs the homogeneous B-S equation whose solution is the running mass.  Section \ref{step3} and Appendix \ref{vpapp} introduce an approximation to the output 3-gluon vertex, based on IR confinement analogs found in IR confining scalar theories in d=5,6.  This section also proposes, for heuristic purposes, a d=3 running charge 
that is just another name for part of the 3-gluon vertex, similar to what has been done in d=4 \cite{corn144,corn147}. Section \ref{landau} estimates the function $1+\widehat{G}(p)$ that determines the ratio between the Landau-gauge lattice propagator $\Delta_L$ and the PT propagator $\widehat{\Delta}$.  Knowledge of $1+\widehat{G}(p)$ allows us to compare the lattice-gauge Landau propagator to the PT propagator, and in particular the gluon masses (defined from the inverse propagators at zero momentum).  The  Landau-gauge mass $m_L$  need not, and does not, agree with the PT mass $m$.   Section \ref{conc} has a summary and conclusions, and   Appendices A, B, C elaborate on the vertex paradigm, scalar theories in d=5,6, and a useful regulator for certain divergent integrals, respectively.

\section{\label{crit}  Critical properties}

Some properties that complicate the   gluon mass problem hold in both the PT Green's functions and in the Landau-gauge lattice gluon propagator, and in both d=3 and d=4.   The Green's functions   have the properties that: 
\begin{enumerate}
\item The inverse propagator has zero-mass scalar poles, akin to Nambu-Goldstone  poles.  These are required by gauge invariance if the proper self-energy does not vanish at zero momentum. 
This non-vanishing is equivalent to a dynamical gluon mass, although not  a simple pole mass.   These NG poles cannot appear in physical quantities.
\item  The propagator, although it obeys a spectral representation, does not have a strictly non-negative spectral function.  This is the positivity problem.  It is a d=3 avatar of AF.
\item The PT 3-gluon vertex has features related to the PT propagator non-positivity through the QED-like Ward identity connecting them.
\end{enumerate}

 In an $R_{\xi}$ gauge the inverse of the PT propagator has the form:
\begin{equation}
\label{loopprop}  \widehat{\Delta}^{-1}_{ij}(p)=P_{ij}(p)\widehat{\Delta}^{-1}(p)+\frac{1}{\xi} p_i p_j,\;\widehat{\Delta}^{-1}(p)=p^2+\widehat{\Pi}(p)
\end{equation}
where the transverse projector is
\begin{equation}
\label{transproj}
P_{ij}(p)=\delta_{ij}-\frac{p_ip_j}{p^2}.
\end{equation}
 The PT proper self-energy $ \widehat{\Pi}(p) $ is independent of the gauge-fixing parameter $\xi$, whose coefficient in the propagator receives no physical corrections.  We omit writing this gauge-fixing term   in the following equations.  

\subsection{First critical property:  A gluon mass}

From Eq.~(\ref{loopprop}) one sees that the first critical property, the NG-like poles, arises as long as  $\widehat{\Pi}(p=0)\neq 0$.   This is equivalent to dynamical mass generation, which is signaled by an inverse propagator that is finite and positive at zero momentum.  Through the QED-like Ward identity relating the 3-gluon PT vertex to the inverse PT propagator, these NG-like poles have to be in the vertex, but these poles get projected out in Landau-gauge lattice studies. 
The second critical property shows up in the lattice  data cited above, most of it in Landau gauge, with clear evidence both of non-positivity in the propagator spectral function.

\begin{figure}
\begin{center}
\includegraphics[width=4in]{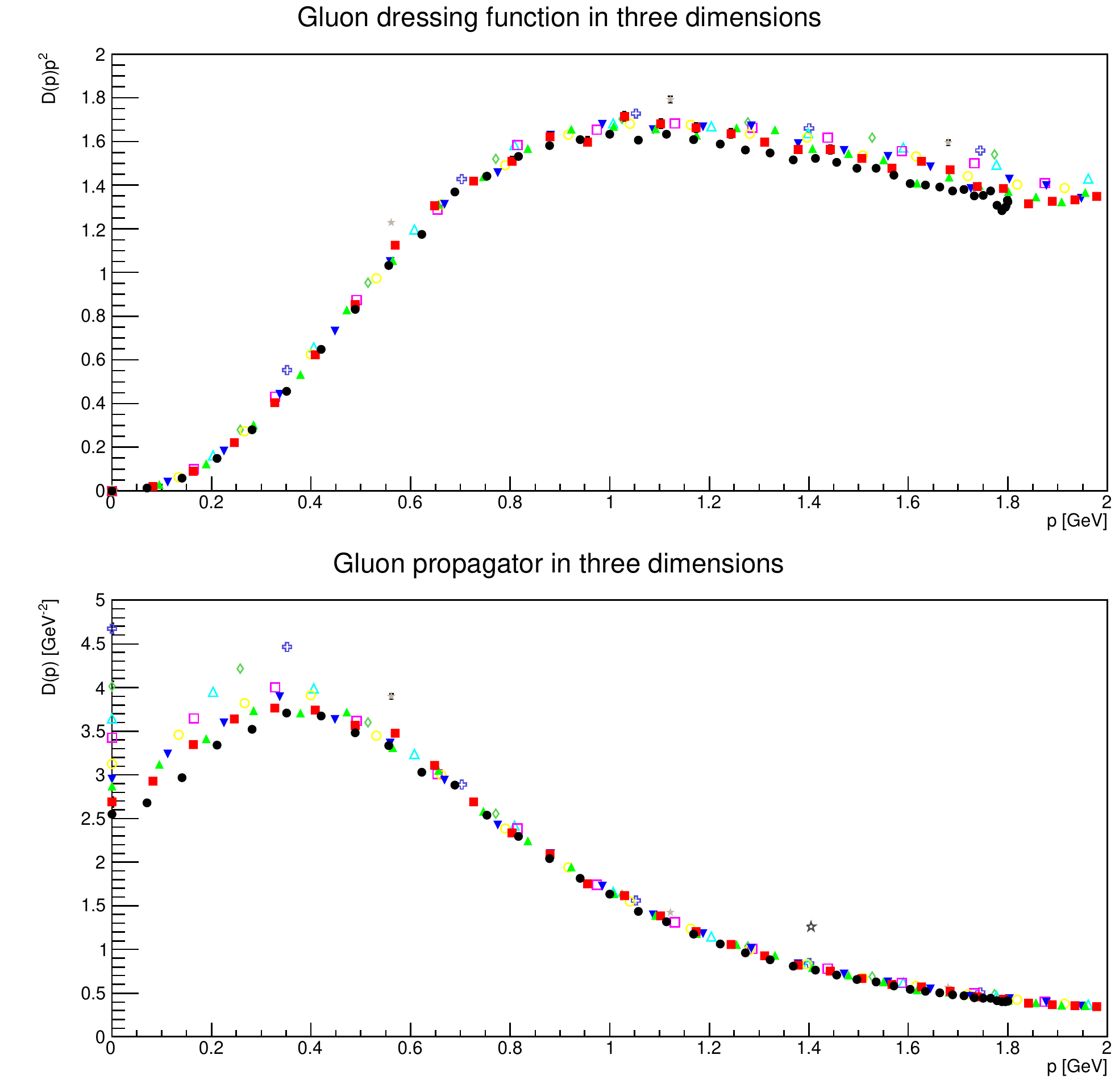}
\caption{\label{maasfig} The d=3 Landau-gauge gluon propagator ($D(p)$; lower curve) as a function of momentum $p$ for various lattice sizes.  The filled circles are at the largest lattice size of 18 fm$^3$.  (The gluon dressing function in the upper graph is $p^2$ times the lower curve.)}
\end{center}
\end{figure}

The first property, a gluon mass, is evident for the Landau gauge in Figure \ref{maasfig},
 showing the d=3 gluon propagator in the Landau gauge \cite{maas}.

\subsection{Second critical property:  Non-positivity}

The second critical property is that the spectral function is negative in some regions.
The scalar function $\widehat{\Delta}$ has the spectral representation   
\begin{equation}
\label{spectrep}
\widehat{\Delta}(p) =\frac{1}{\pi}\int_{\sigma_0}^{\infty}\/d\sigma \frac{\rho (\sigma)}{p^2+\sigma}.
\end{equation}
The same basic representation holds in, for example, the Landau gauge, but unlike the spectral representation for conventional gauge-dependent propagators, in the PT case there are no unphysical and gauge-dependent threshholds, such as would apply to ghosts, and $\sigma_0$ is strictly positive.  If the spectral function is nowhere negative, it is apparent that the derivative with respect to $p$ of the propagator can nowhere be positive, and equally apparent that this condition is violated in Figure \ref{maasfig}.  The filled circles, data for the largest lattice, as well as data for smaller lattices, clearly show that the gluon propagator has a positive slope at zero momentum, which equally clearly shows that positivity is violated.

The  positivity violation in d=4  is not obvious just from a casual glance at the propagator.   A little closer look shows that there is indeed non-positivity in d=4, but not as pronounced as for d=3. See, for example, Fig.~1 of \cite{cucchmen}, comparing the two cases.  In both d=3 and d=4 the cause of the bulge is ``wrong" signs coming from IR confinement.
     
Finding the spectral function itself from lattice data is not straightforward, because  only data in the Euclidean region are available and it is difficult to reconstruct the spectral function accurately just from knowledge of the propagator in this regime.  For a brief review of these issues with references to original work, see \cite{corn145}.

The source of the bulge in d=3,4 is the IR confinement ``wrong" sign. As has long been known \cite{corna9,corn76,alexnair}  the d=3 $SU(N)$ PT propagator in one-loop perturbation theory is:
\begin{equation}
\label{1loop}
\widehat{\Delta}^{-1}(p)=p^2+\widehat{\Pi}(p) = p^2-\pi bg^2p,
\end{equation}
where $g^2$ is the d=3 gauge coupling and $b$ is the {\em gauge-invariant} number
\begin{equation}
\label{gi1loop}
b=\frac{15N}{32\pi}.
\end{equation}
The minus sign in (\ref{1loop}) comes directly from d=4 AF.
In Landau gauge 15 is replaced by 11; this suggests the degree to which the Landau-gauge and the PT propagator differ, although both have the same so-called ``wrong" sign.
It is this sign in this inverse propagator that gives rise to an unphysical tachyonic pole in the perturbative PT propagator at a Euclidean momentum $p^2=(\pi bg^2)^2$.
This tachyon will be killed by a mass term and by massive internal gluon propagators,  provided that the mass is large enough.  But removing the tachyon is not enough; it leaves its mark behind in the  propagator bulge of Fig.~\ref{maasfig}       that unambiguously reveals the non-positivity problem. This means that the output propagator does not resemble a free massive propagator such as in Eq.~(\ref{treeprop}) below except in the extreme IR and UV.   

\subsection{Third critical property:  An inverse bulge in the 3-gluon vertex}

There are no   lattice data that are useful in understanding the PT 3-gluon vertex, but this vertex has critical properties in the gluon gap equation. 
The Ward identity of Eq.(\ref{qedlike}) below, relating the divergence of the 3-gluon Green's function to the inverse PT propagator, suggests that the positivity-violation bulge in the propagator is   mitigated in dressed-loop graphs by an offsetting dip in the vertex.  In the gap equation this can lead to substantial cancellation of non-positivity effects between vertex and propagator.

This is a difficult issue to explore, since the output 3-gluon vertex of the vertex paradigm is so complicated, and has ``wrong"-sign problems of its own that could lead to unphysical tachyons.  At the moment we can only address it heuristically and approximately.  One element is to exploit AF in the scalar theory $\phi^3_6$, having strong analogs to d=4 NAGTs, and its avatar in $\phi^3_5$, analogous to a d=3 NAGT \cite{corn144,corn147}.  In particular, the propagator bulge can be well-modeled, and it is related to a 3-gluon vertex form factor by a QED-like Ward identity, just as in an NAGT.  This suggests that our model for the form factor is also useful, since it is from this form factor and the Ward identity that the propagator is derived. 

In d=4, the 3-gluon form factor of interest is closely related to the usual running charge, which rises from the UV and saturates in the IR.
We hope to make plausible here  that in d=3 there is also a  running charge $\bar{g}(p)$ with the same properties, defined not through a renormalization group but through the PT 3-gluon vertex of the NAGT and the Ward identity relating it to the PT propagator. See Appendix \ref{d3runch} for a few details. Because the renormalization group is not involved, this charge is defined in d=3 as well as in d=4 (where it agrees, through two loops, with the usual running charge in the UV).  We define the running charge by   one of the 3-gluon vertex scalar form factors with one momentum set to zero \cite{corn144,corn147}; let us call this $G(p,-p,0)$.   The running charge $\bar{g}^2(p)$ is related to the form factor   by Eq.~(\ref{runchnum2}), repeated here for convenience.
\begin{equation}
\label{runchnum3}
\bar{g}^{2}(p)=\frac{g^2}{G(p,-p,0)}
\end{equation}  
In the tweaked analog models we identify this form factor with the primary form factor of the Abelian current introduced above.

In the IR it is not possible to define a running charge uniquely, in d=3 or in d=4.  But our definition is physically plausible, and has useful properties.  For example, we will take it that  the squared running charge is, as its name suggest, positive.  Moreover, our definition in  d=3  yields a running charge largest at zero momentum and monotonically decreasing toward the UV, where it approaches the fixed Lagrangian coupling $g^2$ at infinite momentum.  The consequent properties for $G(p,-p,0)$ imply a vertex-function dip that tends to offset the  non-positivity bulge in the gluon propagator.    
The final result is co-existence of the non-positivity bulge in the PT (or Landau gauge) propagator with a gluon mass that is consistent both with the gap equation  and with Landau-gauge lattice data.

\section{\label{vp1} The vertex paradigm, the Ward identity, and   massless scalar poles in the PT propagator} 

We give here only an outline   of the technically-tedious steps in the one-loop vertex paradigm. Appendix \ref{review} gives a very brief summary, and details are in \cite{corn147}.

 The main point of the vertex paradigm, a truncation of the PT SDEs, is to construct successive approximations to the PT gluon proper self-energy and to the PT 3-gluon vertex, following PT principles for constructing gauge-invariant Green's functions, that 
\begin{enumerate}
\item Satisfy the QED-like (ghost-free) Ward identities of the PT
\item  Incorporate dynamical gluon mass in the IR
\item  Yield the correct    perturbative results in the UV
\end{enumerate}
The potential advantage of the vertex paradigm, compared to other truncations the author knows about, is that, in principle at least, it yields not only a plausible semi-quantitative candidate for the PT propagator but also for the PT 3-gluon vertex.  There is a strong connection between these Green's functions from the Ward identity relating them.

There are several obstacles to implementing the vertex paradigm:
\begin{enumerate}
\item Not every approximation to a 3-vertex will satisfy the Ward identity  (see Eq.~(\ref{qedlike}) below) structurally, that is, have a divergence that actually is the difference of two identical functions with different momenta as arguments.
\item A gluon mass requires poles in the inverse propagator and so, by the Ward identity of Eq.~(\ref{qedlike}), also in the 3-vertex.  But no such poles occur in a one-loop vertex constructed with simple input propagators and vertices.
\item The method of successive approximations may show signs of non-convergence.
\end{enumerate}
The vertex paradigm \cite{corn144,corn147} can handle the first two problems.   We devote much of this paper to formulating a semi-quantitative solution to  the last problem, which arises because of non-positivity.  The trouble is that in d=3 the one-loop output propagator does not at all resemble the input propagator.

In the PT, Ward identities are QED-like, with no ghost contributions.  For example, the Ward identity relating the PT 3-gluon vertex $\widehat{\Gamma}_{ijk}$ to the PT inverse propagator is:
\begin{equation}
\label{qedlike}
p_{1i}\widehat{\Gamma}_{ijk}(p_1,p_2,p_3)=\widehat{\Delta}^{-1}(p_2)P_{jk}(p_2)-\widehat{\Delta}^{-1}(p_3)P_{jk}(p_3). 
\end{equation}
Although this is a ghost-free and gauge-independent relation, the right-hand side is not a difference of inverse propagators (except in a ghost-free gauge).  If $\widehat{\Delta}^{-1}(p=0)$ is not zero, there are poles in the right-hand side, and therefore such poles also exist in the vertex.   

In the vertex paradigm, the vertex in the Ward identity    is a sum of two pieces. The first vertex part is a simple Feynman integral and it does not have the poles required by the Ward identity to yield the poles of the massive inverse propagator in Eq.~(\ref{loopprop}).  So we add (see Sec.~\ref{vertform}) a second vertex part \cite{corn90}, called $V_{ijk}$,  which is the product of regular factors times terms with massless longitudinally-coupled NG-like scalar excitations. It satisfies its own Ward identity that gives precisely the pole parts of the inverse propagator.  We emphasize that these NG-like excitations do not imply symmetry breaking in dynamical gluon mass generation for an NAGT.

If we knew the vertex we could find the inverse propagator from the Ward identity.  This seems like a circular statement since one needs the propagator to find the vertex.  We try to avoid this circularity by using successive approximations, starting with  a reasonable tree-level form for input propagators   and vertices in one-dressed-loop graphs.   The approximate output vertex is just the integral over the input propagator and vertices.  Then the Ward identity gives the output inverse propagator.  The hope is that this process of successive approximations will eventually converge.

An earlier truncation method, called the gauge technique and explained in \cite{cornbinpap}, attempts the inverse problem:  Given the propagator, find the vertex.  This is a much easier problem with an algebraic solution (see \cite{cornbinpap} and Sec.~\ref{vertform}), but it is not very accurate.  It leads to SDEs written entirely in terms of the propagator.  The gauge technique is based on the construction of Sec.~\ref{vertform} below.

 As is by now well-known, the PT algorithm is equivalent to working in the background-field method Feynman gauge \cite{cornbinpap}. The first vertex piece, called $G_{ijk}$, is based on the PT or BFM-Feynman approach \cite{ corn99,brodbing,ahmad} to the one-loop 3-gluon vertex in  perturbation theory, with massless gluons and ghosts.    This one-loop vertex is quite complex and has never been evaluated fully, even in perturbation theory.  Fortunately, its graphical construction makes it straightforward to verify the Ward identity from the momentum-space integrand of the one-loop vertex, without actually doing any integrals,  and it is then easy to evaluate the vertex paradigm PT proper self-energy.   

In order to use the mass gap equation we need a semi-quantitative approximation to this complicated output vertex.  The vertex has   some properties that are general consequences of IR confinement and a QED-Ward identity, so we will   model these properties in one-loop graphs of $\phi^3_5$ (see Appendix \ref{vpapp}).   This can be at best semi-quantitatively correct, but it serves to make the point about how non-positivity is subdued by cancellations between the product of propagator and vertex occurring in the mass-gap equation.  

\subsection{\label{vertform} The PT Ward identity and Nambu-Goldstone  poles}
 
The problem of finding $V_{ijk}$ was solved in principle long ago \cite{corn90}.   The full vertex is the sum of these two parts:
\begin{equation}
\label{vertsum}
\widehat{\Gamma}_{ijk}(q,k_1,k_2)= G_{ijk}(q,k_1,k_2)+V_{ijk}(q,k_1,k_2).
\end{equation}
The vertex $V_{ijk}$ has the form, in d=3:
 \begin{equation}
\label{corn090vert}
V_{ijk}(p_1,p_2,p_3)= \frac{-p_{1i}p_{2j}}{2p_1^2p_2^2}(p_1-p_2)_a
\Pi^m_{ak}(p_3)-\frac{p_{3k}}{p_3^2}[P_{ai}(p_1)\Pi^m_{aj}(p_2)-
P_{aj}(p_2)\Pi^m_{ai}(p_1)]+c.p.
\end{equation}
 We have expressed this vertex in terms of a special transverse self-energy $\Pi^m_{ij}$ that is  purely non-perturbative, and define its scalar part by
\begin{equation}
\label{vscalar}
\Pi^m_{ij}(p)=P_{ij}(p)\Pi^m(p)\equiv P_{ij}(p)m^2(p)
\end{equation} 
where $m(p)$ is the running mass.  (In the earliest  PT papers, this self-energy was assumed to be the full self-energy, yielding the gauge-technique truncation of the SDE.)   In the constant-mass approximation we define
$m = m(p=0)$. 
Note that every term in $V$ has not one, but two, massless scalar poles, but its Ward identity has only a single pole:
\begin{equation}
\label{vwi}
p_{1i}V_{ijk}(p_1,p_2,p_3)= \Pi^m(p_2)\frac{p_{2j}p_{2k}}{p_2^2}-\Pi^m(p_3)\frac{p_{3j}p_{3k}}{p_3^2}.  
\end{equation}

 Observe that the Ward identity for the $V$ vertex exactly satisfies the Ward identity necessary to accommodate the poles of the full inverse propagator.  It follows that $G_{ijk}$ must also obey that Ward identity, but with an inverse propagator $\Delta^{-1}$ that has no poles.  The full inverse propagator is the sum of this pole-free part plus the pole terms of in Eq.~(\ref{vwi}).

\subsection{The pole-free part of the Ward identity }

We now come to the hard part \cite{corn147}: To find an approximate and fairly simple form of the pole-free vertex and inverse propagator that exactly satisfies the same QED-like Ward identity:
\begin{equation}
\label{gluewi}   
p_{1i}G_{ijk}(p_1,p_2,p_3)=\Delta^{-1}(p_2)P_{jk}(p_2)-\Delta^{-1}(p_3)P_{jk}(p_3)  
\end{equation}
but with no poles either in the vertex or in the inverse propagator.  
Here the inverse propagator $\Delta^{-1}$ is not the full inverse PT propagator $\widehat{\Delta}^{-1}$, but only that part of it vanishing at zero momentum, thus yielding no poles.  Similarly, $G_{ijk}$ is not the full PT vertex, but its sum with $V_{ijk}$ is the full vertex, as in Eq.~(\ref{vertsum}), and the full inverse propagator is the sum of $\Delta^{-1}$ and the pole terms as in Eq.~(\ref{fullfunction}) below.
In d=4 an approximate   one-loop pole-free 3-gluon vertex   that exactly satisfies the Ward identity is given in \cite{corn147}, with a construction based on a reasonably straightforward, if complicated in detail, extension of the one-loop perturbative 3-gluon PT vertex \cite{corn99}.  IR singularities are removed with free massive propagators as   in Eq.~(\ref{treeprop}) below.  Even in perturbation theory it would be a formidable job to do the integrals for the output three-gluon vertex explicitly \cite{corn99,brodbing,ahmad}, but in order to get the output PT propagator we need only use the QED-like Ward identity of Eq.~(\ref{gluewi}) and the unintegrated vertex.
Now we can add $G_{ijk}$ to the pole vertex $V_{ijk}$, and similarly the  pole part of the self-energy $\Pi^m$ to the self-energy $\Pi$ and form an approximation to the full PT inverse propagator:
\begin{equation}
\label{fullfunction}
\widehat{\Delta}^{-1}(p)\equiv p^2+\widehat{\Pi}(p^2)=p^2+\Pi (p^2)+\Pi^m(p^2).
\end{equation}
The vertex sum of Eq.~(\ref{vertsum})
obeys the full PT Ward identity.

\section{\label{step1}  The d=3 vertex paradigm---Step 1:  The pole-free vertex}

\subsection{Inputs}

The inputs for constructing the pole-free output Green's functions are essentially free massive propagators for gluons and ghosts:
\begin{equation}
\label{treeprop}
\widehat{\Delta}^0_{ij}(p)=\frac{\delta_{ij}}{p^2+m^2};\;\;\widehat{\Delta}_{gh}=\frac{1}{p^2+m^2}.
\end{equation}
plus free vertices.  Their motivation (in particular, why the input ghost has the same mass as the gluon) and  use in the PT are discussed a little further in Appendix \ref{review}.
   These inputs yield outputs   which (besides enforcing both gauge invariance and RGI)    are one-loop exact in the UV, are IR finite, and exactly satisfy the necessary Ward identity, as long as the mass is constant.  Unfortunately, these outputs (with the pole terms added, as described in Section \ref{vertform}) do not much resemble the inputs, because the output propagator has a distinct bulge and the input propagator does not.

\subsection{Pole-free outputs}

We will not give the tedious algebra \cite{corn147} needed to find  the  pole-free 3-vertex and inverse propagator.  They follow from a straightforward adaptation of the d=4 vertex paradigm results \cite{corn147} to d=3.   The result for the full inverse PT propagator including the pole terms is: 
\begin{eqnarray}
\label{3vpprop}
\widehat{\Delta}^{-1}_{ij}(p) & \equiv & \Delta^{-1}_{ij}(p)+P_{ij}(p)\Pi^m(p) = P_{ij}(p)p^2 -\\ \nonumber 
& - & \frac{Ng^2}{(2\pi)^3}\int\,\frac{d^3k}{(k^2+m^2)[(p+k)^2+m^2]}[P_{ij}(p)(4p^2+m^2)+\frac{1}{2}(2k+p)_i(2k+p)_j]  \\ \nonumber
& + &  \frac{Ng^2}{2(2\pi)^3}\delta_{ij}\int\,\frac{d^3k}{k^2+m^2} +P_{ij}(p)\Pi^m(p)\\
\end{eqnarray}
where we have omitted irrelevant gauge-fixing terms, and $P_{ij}(p)$ is the transverse projector.   This is the sum, as in Eq.~(\ref{fullfunction}), of a pole-free term $\Delta^{-1}$ and the new term with $\Pi^m$.  Only this new term contributes to the pole parts of the vertex and inverse propagator.

In the next section we will deal with the determination of the  mass function $\Pi^m(p)$
   multiplying the massless scalar poles (see Eq.~(\ref{corn090vert})) that are necessary for generating a gluon mass in NAGTs.     This term is the analog, in  an NAGT, of the gluonic self-energy part that couples  to the massless scalar excitations in a simple Abelian model of dynamical gluon mass generation given long ago \cite{corn44}.       The massless pole in the $\Pi^m$  term comes  from the $p_ip_j/p^2$ term in the transverse projector.   By gauge invariance $\Pi^m$ also occurs in the non-pole terms, to complete the transverse projector.

For the first evaluation of the vertex-paradigm output propagator we use a constant mass $m$ everywhere, including in the seagull term.
It turns out that for constant $m$  the explicit seagull and the term with numerator $\sim (2k+p)_i(2k+p)_j$  in Eq.~(\ref{3vpprop}) add up to a term that is  transverse and vanishes at zero momentum, although each separately is  non-transverse and non-vanishing at zero momentum.  The reason that a $p$-independent seagull restores transversality to a $p$-dependent integral is the that the divergence of the integral only depends on the value of the integral at $p=0$, as one easily checks by taking the divergence of the term with this numerator.  The calculations require the       regulator formula  of Eq.~(\ref{regrule}) in Appendix \ref{regsec}.  
 Applied to the usual seagull integral---with a constant mass---the regulator yields: 
\begin{equation}
\label{seareg}
\frac{1}{(2\pi )^3}\int\,\frac{d^3k}{k^2+m^2}\rightarrow -\frac{2}{(2\pi )^3}\int\,\frac{d^3k\,m^2}{(k^2+m^2)^2}=-\frac{m}{4\pi}.
\end{equation}
The regulator replaces an integral with a single propagator by an integral with two propagators, which is why the seagull can cancel out the (zero-momentum part of) another term that has two propagators.
In d=4 the regulated integral still diverges (logarithmically) for constant mass, and there is no useful regulation.  The logarithmic divergence can only be removed if the   dynamical gluon mass vanishes at infinite momentum in d=4.   But this is not required in d=3, and the same techniques that in d=3 allow us to estimate quantitatively  the gluon mass  only lead to lower bounds in d=4 \cite{corn147}. 

Again, all these considerations would be much more complicated with a running mass.

\subsection{First step in evaluating the vertex-paradigm output propagator}

From now on we take the pole mass function $\Pi^m(p^2)$ to have the constant value $m^2\equiv \Pi^m(p=0)$. 
Then the basic scalar integral has the value
\begin{equation}
\label{3dint}
\frac{1}{(2\pi )^3}\int\,\frac{d^3k}{(k^2+m^2)[(p+k)^2+m^2]}=\frac{1}{4\pi p}\arctan (\frac{p}{2m}).
\end{equation}
 The remaining steps in the vertex paradigm lead to a self-consistent value for $m^2$.

With these remarks, we write the  output inverse propagator as:
\begin{equation}
\label{invprop}
\widehat{\Delta}^{-1}(p) = p^2  - 2bg^2p\arctan [\frac{p}{2m}]+\frac{8bg^2m^2}{15p}\arctan [\frac{p}{2m}] - \frac{4bg^2m}{15}+m^2.
\end{equation} 
If one drops the final $m^2$ (that is, the $\Pi^m(p=0)$ term) the remainder of this inverse propagator vanishes at zero momentum.
In these equations, $b$ is the parameter of Eq.~(\ref{gi1loop}).

The next step is to find the dynamical B-S equation that governs both the value of the mass, in terms of $g^2$, and the existence of the NG-like massless scalars.  The self-consistency requirement of the vertex paradigm is that the output mass in Eq.(\ref{invprop}) equals the input mass, and the consequences of this consistency condition come from the B-S equation.

\section{\label{mix} The d=3 vertex paradigm---Step 2:  Finding the gap equation} 

For gauge-invariant gluon mass generation there must be an adjoint multiplet of massless scalars, derivatively-coupled to the gluons.  These are the NG-like particles spoken of earlier. They furnish the poles in the vertex and inverse propagator.  

\subsection{\label{step2}  From the mixing amplitude to the gluon propagator}

The effective action
\begin{equation}
\label{mixaction}
S_{mix} = \int \, d^3x\, mTr(A_i\partial_i\phi )
\end{equation}
describes the mixing.
Here the coupling mass $m$ is  the running dynamical gluon mass at zero momentum, and $\phi$ is the composite NG field.    

Self-consistency of the successive approximation scheme requires that
we define the zero-momentum value of the inverse propagator to be the same as the input squared mass $m^2$:
\begin{equation}
\label{zeromomprop}
\widehat{\Delta}^{-1}(p=0)=\Pi^m(p=0)\equiv m^2.
\end{equation}
   (One could then consider $\Pi^m(p)-\Pi^m(0)$ as part of the pole-free self-energy, but we will keep $\Pi^m(p)$ as a separate entity.)  This self-energy comes from a strictly non-perturbative amplitude that mixes the longitudinal part of the gluon with the NG-like particle.

How does the mixing amplitude enter into the gluon propagator?  It must generate the pole term in the inverse propagator, of the form:
\begin{equation}
\label{invproppole}
\widehat{\Delta}^{-1}_{ij}(p)=-m^2\frac{p_ip_j}{p^2}+\dots
\end{equation}
A few minutes' play with Feynman diagrams shows that if two particles A, B have a linear mixing term such as the action of Eq.~(\ref{mixaction}) with strength $\lambda$, the AA inverse propagator has the form
\begin{equation}
\label{2mixprop}
D^{-1}_{AA}(p)=p^2+\Pi_{A;1PI}-\lambda D_{BB} \lambda
\end{equation}
where $\Pi_{A;1PI}$ is the A proper self-energy that is one-particle irreducible (1PI) with respect to A, and $D_{BB}$ is the B propagator that is 1PI with respect to A.  For dynamic gluon mass generation there is no term  $S_{mix}$ in the original action and $\lambda$ should be replaced by a B-S form with one or more loops, as in Fig.~\ref{gluon-mix-fig} below.  Take particle A to be the gluon and B to be the NG boson; comparison of Eqs.~(\ref{invproppole},\ref{2mixprop}) then shows how the mixing amplitude enters.

\subsection{The mixing amplitude}

\begin{figure}
\begin{center}
\includegraphics[width=4in]{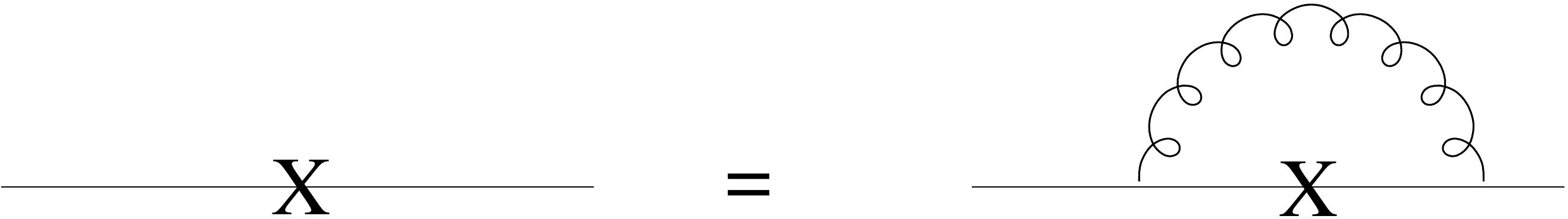}
\caption{\label{gap-eqn-fig} A standard gap equation for the running mass of a quark with CSB.  The cross indicates the insertion of the running mass into the propagator.}
\end{center}
\end{figure}

This amplitude  obeys a homogeneous gap equation, much like the equation for a quark constituent mass coming from chiral symmetry breakdown (CSB)  as shown in Fig.~\ref{gap-eqn-fig}.
If there is a solution to this homogeneous equation, then there is CSB and spontaneous fermion mass generation.  But at the same time, the gap equation is the zero-momentum Bethe-Salpeter equation for a massless triplet of pions, so the Nambu-Goldstone mechanism works for composite NG bosons.   

There are some critical differences for dynamical gluon mass generation.  First, there is no symmetry being broken, and second, the gap equation refers to a mixing amplitude between particles of very different character:  The gluon and the NG particle.  However, although not usually thought of as such,  a dynamical mass for a chirally-symmetric quark is a mixing process between different particles, coupling left-handed and right-handed quarks.  In the case of NAGT gluons, the mixing process  adds the third longitudinal polarization state needed for a massive gluon.

\begin{figure}
\begin{center}
\includegraphics[width=4in]{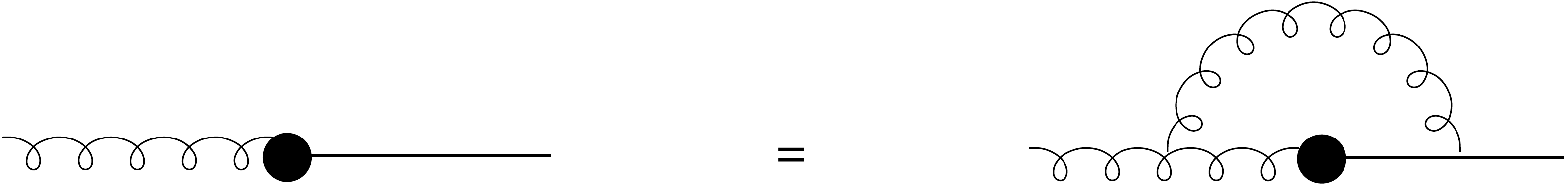}
\caption{\label{gluon-mix-fig}  The dynamical gluon mass equation for the mixing amplitude described by the action in Eq.~(\ref{mixaction}), and indicated by the black circles.}
\end{center}
\end{figure}

The steps to follow are familiar indeed in related contexts, but technically more involved because of the proliferation of spin indices on gluons and the fact (see \cite{corn44} and references therein) that the scalar pole, just like an NG particle, cannot occur in a physical amplitude.    Since the NG particle is in the adjoint representation, it has a standard gauge coupling to the gauge boson, with strength $g$.  Elementary symmetry considerations show that there is no coupling of one NG particle to two gluons.  In consequence, the one-loop B-S equation describing the NG field has the graphical representation of Fig.~\ref{gluon-mix-fig}.
There is also a seagull graph that enforces gauge invariance, which we do not show.

At first glance, this equation seems to violate, because of the massless internal line in the figure, the well-known principle that NG particles cannot occur in the S-matrix or other physical quantities, such as the running mass.   But in fact there is no pole for this line, because of a cancellation brought about in the numerator.  In Fig.~\ref{gluon-mix-fig} let $q$ be the momentum of the internal NG line and $p_i$ be the external momentum. Then the graph in the figure has a kinematic factor of $p_i$ multiplying a scalar graph.  The momentum dependence of the numerator of the graph comes out to be:
\begin{equation}
\label{graphnum}
-2(p_ip\cdot q-q_ip^2) -2(q^2p_i-q_ip\cdot q).
\end{equation} 
The first term is orthogonal to $p_i$ for all $q$, and hence contributes zero, since the graph itself must be proportional to $p_i$.  The second term is orthogonal to $q_i$ and vanishes for the component of $q$ along $p$.  This suggests, and calculation confirms, that the second term in this numerator can be replaced by
\begin{equation}
\label{newnum}
-\frac{4}{3}q^2p_i
\end{equation}
since only two of three directions of $q$ can contribute.  Now one sees that the $q^2$ in the numerator cancels the NG pole, and what remains is a one-loop self-energy graph for two scalars of mass $m$, as a function of $p$.  Note that there is always a solution for a running mass $m(p)$ in this equation because $g^2$ has the dimensions of mass.  Of course, the solution may or may not be  reasonably accurate.   
We can find the  leading term in $m(p)$ at large momentum by evaluating this scalar self-energy with constant mass, and it leads to $m^2(p)\sim 1/p^2$, as the operator product expansion dictates \cite{lavelle}. 

As we said in the beginning, one of the virtues of d=3  is that it is possible to find a description of dynamical gluon mass generation with a mass $m$ that does not run, which is the running mass $m(p)$ evaluated at zero momentum.     The simplest equation for $m$ comes from evaluating the B-S equation   at $p=0$, using the input propagators of Eq.~(\ref{treeprop}):
\begin{equation}
\label{bseqn}
1=\frac{4Ng^2}{3(2\pi )^3}\int\,\frac{d^3q}{(q^2+m^2)((p+q)^2+m^2)}|_{p=0}=\frac{Ng^2}{6\pi m}.
\end{equation}
Taken as it stands, this equation yields
\begin{equation}
\label{mass1}
m=\frac{Ng^2}{6\pi},
\end{equation}
which is a factor of 2 or 3 less than in other works.  

In fact, the actual mass ratio $m/g^2$ could possibly be very different because of non-positivity effects in the 3-gluon vertex.  
In Sec.~\ref{step3} below we show that a simple approximation to the 3-gluon vertex  leads to a tachyonic pole in the propagator unless the gluon mass is large enough, and for the same ``wrong"-sign reason leading to non-positivity.  But in Sec.~\ref{disprel} we cure this pole through a dispersion relation.

\section{\label{step3}  The d=3 vertex paradigm---Step 3:  Dressing the gap equation}

\subsection{Dressed propagator}

The next step in understanding the gap equation is to dress the lines and vertices in Eq.~(\ref{bseqn}).  First  we    evaluate the integral in the gap equation  with the output propagator of Eq.~(\ref{invprop}) and bare vertices.    Using a dressed propagator with bare vertices is a common approximation in dealing with gap equations.

\begin{figure}
\begin{center}
\includegraphics[width=4in]{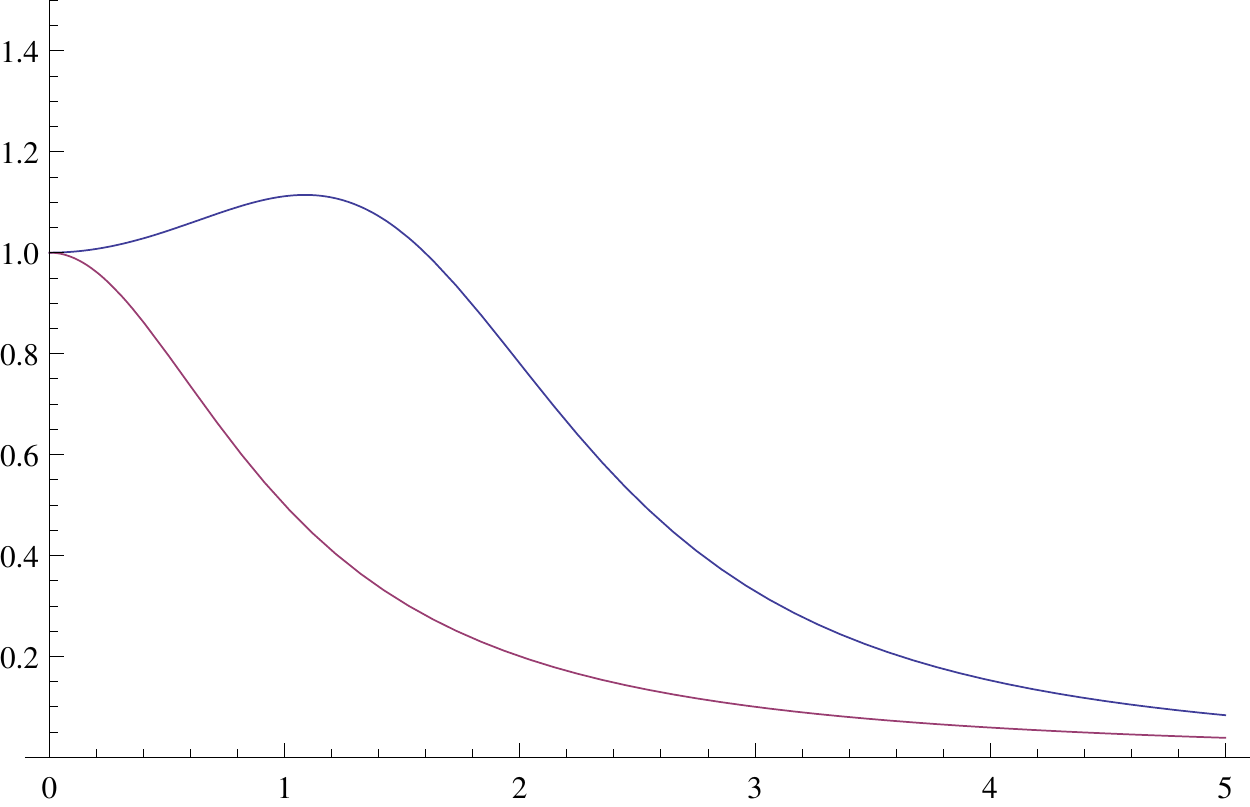}
\caption{\label{3prop}  The lower (red) curve is the free massive propagator, and the upper curve is the vertex paradigm output propagator, for the mass given in the text.}
\end{center}
\end{figure}

Unfortunately, the dressed propagator in Eq.~(\ref{invprop}) is very different from the input propagator.  A comparison of the input propagator and the output propagator, shown in Fig.~\ref{3prop}, is a measure of the potential discrepancy.  This figure makes the comparison  for the specific mass value $m=Ng^2/(2.48 \pi)$ which    is, as we will see, the final estimated value of the self-consistent mass.
Clearly, the output propagator 
does not   resemble the input, because of the bulge in the output propagator.  The bulge is  evident not only in the Landau-gauge propagator, but also in the vertex paradigm output.  It is also clear that the dressed output propagator must give a much larger value to the gap-equation integral than the input propagator will give.
In fact, the  resulting integral  with the dressed propagator is a factor of 6.68 larger than quoted in Eq.~(\ref{bseqn}) and the estimated mass is larger than given in Eq.~(\ref{mass1}) by the same factor.  Such wild swings would not encourage one to believe that one is making progress on the d=3 gluon mass problem by successive approximation methods.   We now show that dressing the 3-gluon vertex substantially mitigates the problem.

\subsection{Approximate dressed vertex}

Our task now is to estimate the effect of non-positivity on  the gap equation both from the propagator and from the vertex.  
One should not make the natural supposition that finding the propagator from the Ward identity for the vertex means that we know   the vertex explicitly.  Given the 3-gluon vertex as a momentum integral over tree-level propagators, it is a much simpler problem  to find the results of a one-loop Ward identity  than it is to evaluate that integral \cite{corn99}.  In fact, that has never been done even in one-loop perturbation theory.  So for this exploratory study of gluon mass with non-positivity we will ultimately come to a heuristic and qualitative form for the 3-gluon vertex that someday must be supplanted by more accurate calculations.

Even without full knowledge of the dressed vertex there is a non-positivity effect in the vertex that ameliorates the effect from the dressed propagator and that can be qualitatively appreciated directly from the Ward identity of Eq.~(\ref{qedlike}):  The bigger the propagator, the smaller the vertex.  The B-S equation (\ref{bseqn}) effectively has the product of two gluon propagators and two 3-gluon vertices in it, and so it is considerably less sensitive to non-positivity effects than either of the pieces is.  We simplify this exploratory study by omitting  the massless pole parts of the vertex since they cannot appear in a physical amplitude, and approximate  the remaining pole-free part as described in Section \ref{disprel} below.   Because we are using the approximation of a non-running mass $m\equiv m(p=0)$, we remove the momentum dependence from the B-S equation by evaluating it at zero momentum.
Then each vertex is the scalar function $G(q,-q,0)$, where $G$ is (an approximation to) the appropriate scalar vertex function.  For want of a better approximation we take these functions from the   model described in Appendix \ref{vpapp}. It is based on an extension of previous work \cite{corn141,corn144,corn147} that models d=4 NAGT effects on the asymptotically-free  scalar theory $\phi^3_6$, and described briefly in Appendix \ref{tweak}. The extension uses a tweaked version of $\phi^3_5$. The approximate (scalar ) B-S equation becomes
\begin{equation}
\label{bseqn2}
1=\frac{4Ng^2}{3(2\pi )^3}\int\,d^3q \,G^2(q,-q,0)\widehat{\Delta}^2(q). 
\end{equation} 
 In this equation the scalar function $G$ is intended to model the scalar form factor of the d=3 3-gluon vertex, also called $G$ and defined in Eq.~(\ref{defg}).  As such, it appears in the Ward identity (\ref{qedlike}). 

 Appendix \ref{vpapp} gives a first approximate form  for this form factor:  
\begin{equation}
\label{gapprox}
 G(p_1,p_2,p_3)=1-2 bg^2\int\,\frac{[dz]}{(D+m^2)^{1/2}}
\end{equation}
where the  factor $(D+m^2)^{-1/2}$ is the denominator of  the d=5 equal-mass scalar triangle graph, as well as the appropriate factor for d=3 NAGTs with massive input propagators.  The negative sign is inherent in $\phi^3_5$, but the value  $2bg^2$ is chosen so that the d=3 Ward identity is valid to $\mathcal{O}(g^2)$ in the massless (large-momentum) limit.  That is, the term of this order in $G$ corresponds to the same term in the output propagator of Eq.~(\ref{1loop}).  
   The Ward identity tells us that $G$ behaves inversely to the propagator $\widehat{\Delta}$, so the propagator bulge ends up being a vertex dip.  The minus sign responsible for the vertex dip expresses the d=3 realization of IR confinement, just as it is responsible for the propagator bulge; see Eq.~(\ref{1loop}).

The vertex $G$ of Eq.~(\ref{gapprox}) is not usable as it stands, because for the self-consistent mass value the vertex has a zero in the Euclidean region.  Appendix \ref{d3runch}   gives a way around this problem, using a dispersion relation for $G^{-1}(p,-p,0)$, and an argument first given \cite{corn144} for the d=4 3-gluon vertex that relates the inverse vertex to the squared running charge $\bar{g}^2(p)$.  The running charge is not defined through a renormalization group or a beta function, and the definition applies to d=3 as well as d=4.     It is not essential for subsequent calculations   that the inverse of a vertex function with one momentum zero is a running charge squared, because  in the end, everything is expressed in terms of $G(p,-p,0)$.  However, the idea that there is a relation such as (\ref{runchnum}) below leads us to insist that $G(p,-p,0)$ be positive.

\subsection{The running charge concept in d=3}

The Ward identity (\ref{qedlike}) provides  a path to a running charge $\bar{g}^2(p)$ that agrees with the PT running charge to two loops at high momenta,  is well-defined and physically-reasonable at all momenta, and does not rely upon a renormalization group or beta-function for its definition.  The last feature makes it possible to use it in d=3.  According to the reasoning of Appendix \ref{d3runch}, this running charge is:
\begin{equation}
\label{runchnum}
\bar{g}^{2}(p)=\frac{g^2}{G(p,-p,0)}.
\end{equation}
 
The above definition equates $G^{-1}(p,-p,0)$ to an ostensibly positive quantity, the square of a running charge.  However, IR confinement interferes.  Based on (\ref{gapprox}) the corresponding    approximation  to $G(p,-p,0)$ would be:
\begin{equation}
\label{gapprox2}
G(p,-p,0)\approx 1-\frac{2bg^2}{p}\arctan (\frac{p}{2m}),
\end{equation}
clearly not positive in general, and as said above, this non-positivity comes from IR confinement in d=3.  If this formula has a zero it completely spoils its interpretation as an inverse running charge, which would not only have a tachyonic pole but would also change sign from positive to negative.

It could happen, but does not, that the self-consistent mass $m$ is so large that $G(p,-p,0)$ is nevertheless positive for all Euclidean momenta.  
Unfortunately, for the self-consistent mass used for Figure \ref{3prop} the approximate $G(p,-p,0)$ does have a Euclidean zero, which the Ward identity translates   into an unwanted tachyonic propagator pole.  While this is not necessarily fatal since the pole need not appear in the S-matrix (because the vertex zero cancels it) it is unnecessary; it also results in in the unphysical result $\bar{g}^2(p)<0$ for a finite range of momentum.   

We proceed to a second step in modeling the vertex dip that removes the zero of the approximate vertex by postulating a dispersion relation for $G^{-1}(p,-p,0)$ (or equivalently the running charge).
\subsection{\label{disprel}  A dispersion relation for the vertex form factor} 

The type of dispersion relation we use here is sometimes invoked  under the name of analytic perturbation theory \cite{shirk}, but our use of it has nothing to do with this subject.

 The building block of the dispersion relation is the simple formula
\begin{equation}
\label{spectral}
\int_{4m^2}^{\infty}\frac{d\sigma}{\sqrt{\sigma}(\sigma + p^2)}=\frac{2}{p}\arctan(\frac{p}{2m}).
\end{equation}
with a manifestly positive spectral function.
Now construct a dispersion relation for $\bar{g}^2(p)$, based on the approximate form of Eq.~(\ref{gapprox2}) and this building block:
\begin{equation}
\label{disprelation}
\bar{g}^2(p)=g^2[1+\frac{1}{\pi}\int_{4m^2}^{\infty}\,\frac{d\sigma}
{\sigma +p^2}Im\,\bar{g}^2(\sigma )]
\end{equation}
with
\begin{equation}
\label{impart}
Im \bar{g}^2(\sigma) =\frac{\pi b g^4}{Q(\sigma )\sqrt{\sigma}}
\end{equation}
and
\begin{equation}
\label{defq}
Q(\sigma )=[1-bg^2\mathcal{P}\int \,\frac{d\sigma'}{\sqrt{\sigma'}(\sigma'- \sigma)}]^2+[\frac{\pi bg^2}{\sqrt{\sigma}}]^2
\end{equation}
There is a subtraction at infinity, corresponding to the appearance of 1 in the equation (\ref{gapprox2}) for $G$, and by hypothesis there is no other subtraction that would yield a pole in the running charge.  Just as in d=4 this running charge is non-negative and monotone decreasing with momentum from a finite positive value at $p=0$.  But in d=3 it approaches the value $g^2$ at infinity.  

With this form for the running charge, we now use the formula (\ref{runchnum}) for $G(p,-p,0)$, inverse to $\bar{g}^2(p)$.  It approaches 1 at infinity and is less than 1 for all finite momenta, but by construction it has no zeroes. Its spectral function is negative, as it must be if it is inverse to a function with a positive spectral function.   The precise analytic expression of this dispersion integral is complicated and unnecessary for our purposes.  As it happens, a simple modification of the original approximate formula (\ref{gapprox2}) is sufficiently accurate   as a standin for the  dispersive integral:
\begin{equation}
\label{analyt}
\frac{g^2}{\bar{g}^2(p)}=G(p,-p,0)=1-\frac{ 0.95bg^2}{p}\arctan(\frac{p}{2m})
\end{equation}
in which the coefficient   is $ 0.95bg^2$ instead of $2bg^2$.  This misstates the vertex in the deep UV, but this is not of concern since mass generation is purely an IR issue and UV contributions are not as important.  For the self-consistent mass there is no Euclidean zero, and it is numerically reasonably close to the dispersive form.  There is no reason to suppose that the coefficient 0.95 is highly accurate;  depending on how the standin vertex is fit to the numerical vertex, this coefficient might change by $\pm 20\%$.

As advertised, the corrected vertex tends to offset the propagator bulge.   At the self-consistent mass  $m=Ng^2/(2.48 \pi)$ it has the zero-momentum value   0.5-0.6, a reduction   below the constant-vertex value of one.  The zero-momentum running charge is inverse to these numbers, that is, between 1.66 and 2.  Just as in a d=4 NAGT the running charge grows in the IR.

\subsection{Final results}

After numerical integration of the formulas of the last section, the Bethe-Salpeter self-consistency relation that replaces the original of Eq.~(\ref{bseqn}) is:
\begin{equation}
\label{numerics}
1=\frac{1.9 \times 2 Ng^2}{3\pi^2 m}
\end{equation}
provided that these integrals are evaluated with the mass value
\begin{equation}
\label{mass2}
m=\frac{Ng^2}{2.48\pi }.
\end{equation}
This is numerically consistent with Eq.~(\ref{numerics}).  The B-S integral has been   enhanced from the bare integral of Eq.~(\ref{bseqn}), but not nearly as much as if bare vertices and the output propagator  (shown in Fig.~\ref{3prop}) were used in the B-S integral.  The reduction comes from the decreasing value of $G$ as the momentum decreases.

As a result,  the mass value coming from the bare equation (\ref{mass1}) of $m=Ng^2/(6\pi )$ is considerably changed.   We might compare to other published mass values by introducing a number $\zeta$, with
\begin{equation}
\label{zetaeq}
m=\frac{Ng^2}{\zeta \pi}.
\end{equation}
   Our present value for $\zeta$ is 2.48.   Various authors have given mass values, but not necessarily the mass as defined by us, as related to the propagator at zero momentum.  Values estimated with gauge-invariant techniques include Ref.~\cite{alexnair}, giving $\zeta = 1.68$; \cite{corn120} gives $\zeta = 2.57$; \cite{nair} has $\zeta = 2.00$; and \cite{agbinpap} claims $\zeta = 2.18$.    The authors of \cite{buchphil,buchphil2} invoke a Higgs field, and Ref.~\cite{corn120} attempts to remove the Higgs mechanism by taking the Higgs mass to infinity, leaving only the NG bosons that occur both in the Higgs mechanism and in the PT.  The result is that   $\zeta = 2.24$ , which in principle would be gauge-invariant.   For whatever it is worth, the average of these numbers is $\zeta = 2.19$, and the spread around the mean is roughly 20\%.   

Although it is gauge-dependent and need not agree with the PT mass, we can  define a Landau-gauge mass $m_L$ and the corresponding $\zeta_L$ as
\begin{equation}
\label{zetalandau}
 \Delta_L(p=0)\equiv 1/m_L^2;\;m_L=(Ng^2)/(\zeta_L\pi).  
\end{equation}
For numerics we take   $  \Delta_L(p)$ as the d=3 Landau-gauge propagator   shown in Fig.~\ref{maasfig}, for which $\zeta_L=2.05$.  Next, we discuss a relation between $m$ and $m_L$ and evaluate their ratio approximately.   

\section{\label{landau} From the PT to the Landau gauge}

If lattice information were available on the PT propagator and vertex, we could stop here.
Unfortunately, such data do not yet exist, but extensive propagator data are available in Landau gauge.  It is, in fact, possible for us to use these data along with Eq.~(\ref{replaceg}) below to give another estimate of the PT mass, not dependent on the gap equation   that yielded $\zeta \approx 2.48$.  To do this, use a background-quantum identity  (used somewhat differently in \cite{agbinpap}), reviewed in \cite{cornbinpap}, that relates the Landau-gauge propagator $\Delta_L$ to the PT propagator $\widehat{\Delta}$:
\begin{equation}
\label{bqi}
\Delta_L(p)=(1+\widehat{G}(p))^2\widehat{\Delta}(p).
\end{equation}
[We have used the simpler notation
\begin{equation}
\label{kappadef}
\kappa_L=[1+\widehat{G}(0)]
\end{equation}
in Eq.~(\ref{bqident}).]

The function $\widehat{G}$ is, in principle, computable in terms of  Landau-gauge Green's functions involving ghosts, but these obey their own SDEs that are not elementary to solve.  Instead, we will use a simple approximation, in the spirit of the approximations already made for the 3-gluon vertex.  The asymptotic UV behavior is easy to find, since it comes from one-loop perturbation theory, and by comparing one-loop results in the PT, given in Eq.~(\ref{1loop}), and in the Landau gauge, we find the UV behavior
\begin{equation}
\label{bqiuv}
1+\widehat{G}(p)\rightarrow 1-\frac{2\pi bg^2}{15 p}.
\end{equation}
Not unexpectedly, this has a forbidden tachyonic pole in the Euclidean regime.  We cure it by the simple expedient of the replacement of the massless perturbative one loop integral by the massive one, and propose:
\begin{equation}
\label{replaceg}
1+\widehat{G}(p)\approx 1-\frac{4bg^2}{15p}\arctan (\frac{p}{2m});\;\;1+\widehat{G}(0)\approx 1-\frac{\zeta}{16}.
\end{equation}
Unlike the original approximation for the 3-gluon vertex of Eq.~(\ref{gapprox}), for the estimated value of $m/g^2$ in Eq.~(\ref{mass2}) this expression is non-singular in the Euclidean regime of real positive $p$  and we will use it as it stands. (Using the dispersion-relation approach as we did for the 3-gluon vertex makes little difference in the Euclidean regime.) Then with $\zeta = 2.48$ we find $1+\widehat{G}(p=0)= 0.845$, and of course $\widehat{G}(p=\infty )=1$.  (At $p=0$  the workers of \cite{agbinpap} replace the 4/15 in Eq.~(\ref{replaceg}) by 32/45.)  The variation of $\widehat{G}(p)$ is so little in the IR that it is not worth plotting; the major IR effect is just rescaling the propagator from the PT to the Landau gauge.  

We use the approximation   in Eq.~(\ref{replaceg})to relate our PT estimates and Landau-gauge lattice data in two ways.   The  rescaling  of Eq.~(\ref{bqi}) leads at zero momentum to the  relation
\begin{equation}
\label{rescale}
m=[1+\widehat{G}(0)]m_L,
\end{equation}
which with the aid of Eq.~(\ref{replaceg}) becomes the quadratic equation
\begin{equation}
\label{quadeq}
\frac{\zeta_L}{\zeta}=1-\frac{\zeta}{16}.
\end{equation}  
If now we use the gap-equation value $\zeta=2.48$ of Eq.~(\ref{numerics}) to solve for $\zeta_L$,  we find $\zeta_L$= 2.10, to be compared to the lattice value of 2.05.  If, on the other hand, we ignore the
gap equation and use the lattice data for $\zeta_L$ to estimate the PT value $\zeta$, the relevant root of the quadratic equation (\ref{quadeq}) is $\zeta$ = 2.41, compared to the gap-equation value of 2.48.  This unnaturally close agreement   can only be a coincidence, given the roughness of our approximations. 
Further and more accurate work is necessary.

 \section{\label{conc}  Summary and conclusions}

Landau-gauge lattice simulations of the gluon propagator in d=3 show a quantitatively-important positivity violation that could be a serious problem for studies of non-perturbative effects such as dynamical gluon mass generation with the gauge-invariant PT.
We study this effect with the vertex paradigm, a method for truncating the PT Schwinger-Dyson equations that is in principle more accurate for the 3-gluon vertex than some other truncation schemes and so far implemented only at the one-dressed-loop level.  We complete the vertex paradigm with a homogeneous Bethe-Salpeter equation describing a set of composite Nambu-Goldstone (NG) bosons that are essential to describe gluon mass generation gauge-invariantly; these NG bosons cancel out of all physical quantities.  However, implementing the vertex paradigm and the B-S equation with successive approximations could lead to serious quantitative errors from non-positivity.  These are substantially mitigated by the fact that in the crucial B-S equation non-positivity effects in the output 3-gluon vertex are in the opposite direction from those in the gluon propagator, as the QED-like Ward identity of the PT shows, and significant cancellation can occur.  

In the successive-approximation scheme, the input ghost and gluon propagators are free propagators with mass $m$, and the input vertices are the usual tree-level ones.   There are two parts to the vertex:  A pole-free part constructed by modification of the perturbative one-loop 3-gluon vertex, and a part containing the NG massless poles.   The part with NG poles algebraically satisfies the Ward identity relating its divergence to the poles of the inverse PT propagator, as in Eq.~(\ref{loopprop}).  The full vertex satisfies the Ward identity of Eq.~(\ref{qedlike}).

It is straightforward, if tedious, to write the momentum-space integral for the output pole-free 3-gluon vertex  but it is far from easy to compute the complete vertex, which in fact has not even been done for the one-loop 3-gluon vertex in perturbation theory. But it is not difficult, given the momentum-space integral, to use the Ward identity to express the inverse propagator as the divergence of the 3-gluon vertex, provided that the mass $m$ is non-running and the same for all gluons.     In view of the difficulty in actually calculating the 3-gluon vertex itself, we approximate it using a tweaked version of $\phi^3_5$.   This scalar theory is a descendant of the asymptotically-free theory $\phi^3_6$ in one higher dimension, and can be used as a heuristic model for  d=3 NAGT phenomena, including non-positivity effects.  From this model plus a dispersion relation that guarantees the absence of unphysical tachyons coming from ``wrong" signs, we construct an approximate model for one of the scalar form factors of the 3-gluon vertex  when one of its momenta vanishes.      As noted in the main text, gluon mass generation can tame the completely unphysical tachyonic pole in the propagator down to a non-positivity bulge.    Absence of these singularities allows us argue (as previously suggested for d=4 NAGTs \cite{corn144,corn147}) that the Ward identity suggests an interpretation for the scalar form factor  as the square of a running charge that is defined without reference to a renormalization group or beta-function.       As the Ward identity (\ref{qedlike}) suggests, a bulge in the propagator results in a dip in the 3-gluon vertex.

  Then we use this approximate 3-gluon vertex in the gap equation in the constant-mass approximation, and note that there is a certain amount of cancellation between the propagator bulge and the vertex dip.  In terms of a parameter $\zeta$ we define the PT mass as $m=Ng^2/(\zeta\pi)$.   With no accounting for non-positivity has the unacceptably large value of 6.  The results of the present paper, taking into account non-positivity and the Ward identity between vertex and propagator, give the lower value $\zeta$ = 2.48.   To compare with published data on the Landau-gauge propagator
 we use a background-quantum identity   telling us that the Landau-gauge propagator and the PT propagator differ at zero momentum by a scale factor $[1+G(0)]^2  < 1$ .   We make a simple estimate of this scale factor,  and use this estimate plus the gap-equation mass value   to estimate the lattice Landau-gauge mass parameter  as $\zeta_L$ =2.05, compared to the simulation value of  2.10.  Conversely, using lattice data and the scale factor, but not the gap equation,   leads to a PT mass parameter estimate of $\zeta$ = 2.41, as compared to the PT estimate of 2.48.  The closeness of the comparisons is undoubtedly fortuitous.

\newpage

\appendix

\section{\label{review}  A brief review of the vertex paradigm}

The following is a road map, not a complete exposition.  The main point is to suggest how to organize matters so that the NG-like massless poles are cancelled as much as possible before carrying out any serious calculations.   This leads us to the pole-free vertex discussed in the main text.

The first step is to construct a tree-level model that has gluon and ghost masses in it.  The action is the usual NAGT action plus a gauged non-linear sigma model mass term, plus gauge-fixing terms.  

Pinching is greatly simplified in the Feynman gauge, which we use for the GNLS model.  In this model the gluon propagator is
\begin{equation}
\label{inputprop2}
\Delta^0_{ij}(p) =\frac{\delta_{ij}}{p^2+m^2} +\frac{p_ip_j}{p^2}[\frac{1}{p^2}-\frac{1}{p^2+m^2}].
\end{equation}
The last term $\sim p_ip_j$, a difference of massless and massive scalar propagators, suggests that the massless ghosts are cancelled out and replaced  by ghosts of mass $m$, as in the Feynman-FLS gauge \cite{fls}, and this is indeed what happens.  Some such replacement must occur in the PT.   In general, the ghost mass is gauge-dependent, and since the PT propagator has only physical threshholds, the massless ghost poles get replaced by poles at $m^2$, in much the same way as it happens in the PT for electroweak theory \cite{cornbinpap}.  Or one may simply argue that the PT propagator can be constructed in a ghost-free gauge, because it is the same in any gauge.  
   So for practical purposes we use the tree-level propagators of Eq.~(\ref{treeprop}), repeated here for convenience,
\begin{equation}
\label{treeprop2}
\widehat{\Delta}^0_{ij}(p)=\frac{\delta_{ij}}{p^2+m^2};\;\;\widehat{\Delta}_{gh}=\frac{1}{p^2+m^2}.
\end{equation}
that have no massless poles.

\begin{figure}
\begin{center}
\includegraphics[width=5in]{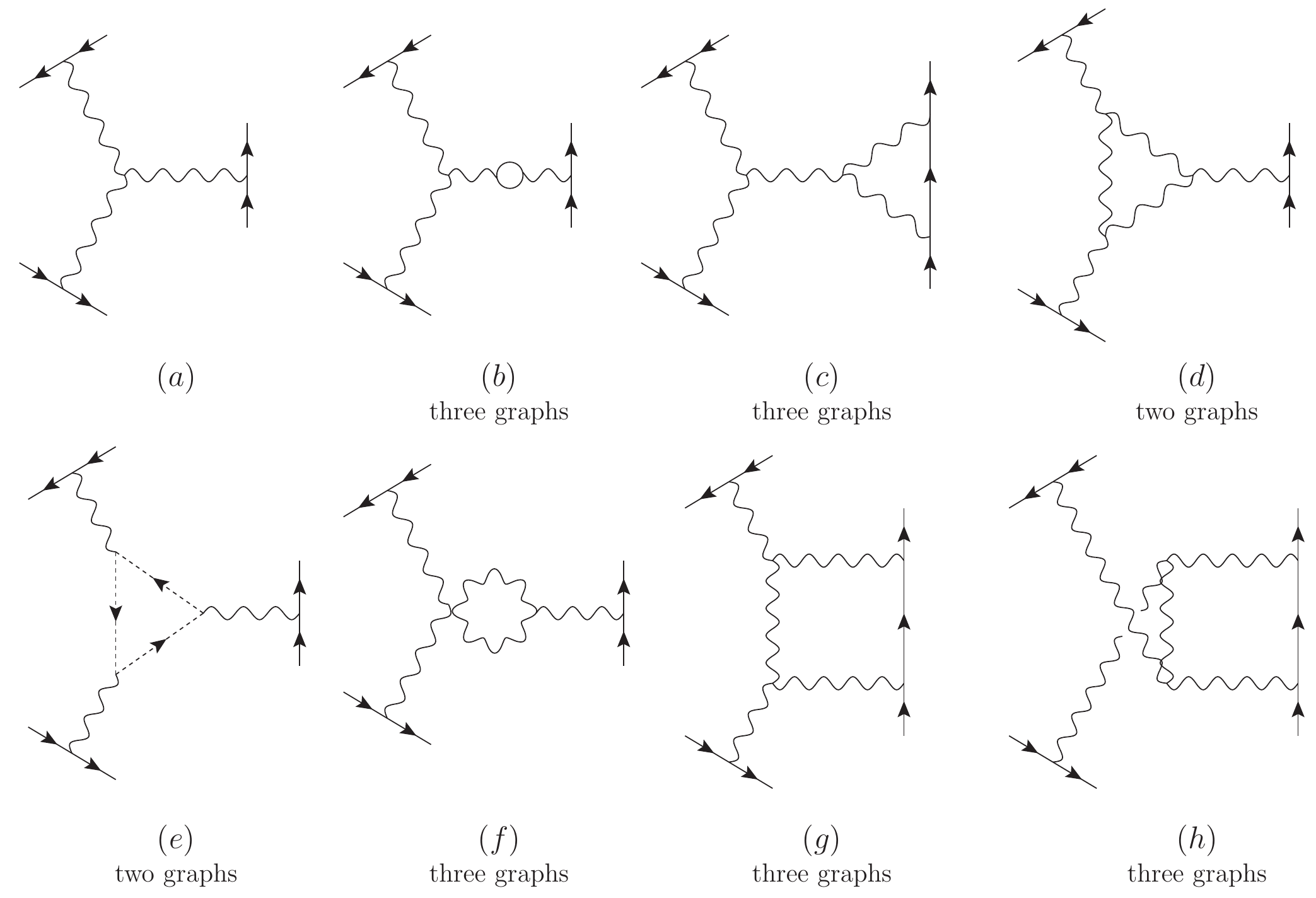}
\caption{\label{vertgraph} The one-loop S-matrix element for finding the PT 3-gluon vertex.  Solid lines represent quarks.}
\end{center}
\end{figure}

The second step is to follow \cite{corn99} and write down the sum of one-loop graphs that give the S-matrix element for the scattering of three external quarks.  Fig.~\ref{vertgraph} shows these graphs.
Of course, some of these graphs, {\em e.g.}, (g),  are  not vertex parts, but they contain vertex parts that are extracted by using tree-level Ward identities that pinch out the internal parts of quark lines.  

Satisfying the PT Ward identity at one-loop level is a matter of satisfying it at tree level.  The form quoted in the text (Eq.~(\ref{qedlike}), although ghost-free, is not really useful at tree level because of its massless poles.  Instead \cite{cornbinpap}, we write the usual tree-level vertex as the sum of two parts:
\begin{equation}
\label{bfmvert}
\Gamma_{ijk}(p_1,p_2,p_3)=\Gamma^F{ijk}(p_1,p_2,p_3)+ \Gamma^P_{ijk}(p_1,p_2,p_3)
\end{equation}
where $\Gamma^F$ is the BFM-Feynman-gauge vertex.  It has one line (called the background line) singled out; say it is $p_1$.  The background lines are those attached directly to quark vertices in Fig.~\ref{vertgraph}.  The remaining part, $\Gamma^P$, has only longitudinal terms $\sim p_{2j},p_{3k}$ that trigger pinch parts.  On the $p_1$ line, $\Gamma^F$ satisfies a simple Ward identity with no massless poles:
\begin{equation}
\label{bfmwi}
p_{1i}\Gamma^F_{ijk}(p_1,p_2,p_3)=  [\widehat{\Delta}^0_{ij}(p_2)]^{-1}-
[\widehat{\Delta}^0_{ij}(p_3)]^{-1}.
\end{equation}
It obeys this Ward identity even for the massive propagator of Eq.~(\ref{treeprop}), provided that the mass $m$ does not run with momentum and is the same for all gluons and ghosts.  It certainly would not satisfy this identity for a running mass.  This is the fundamental reason that we use the constant-mass approximation in the d=3 problem.  
 Satisfying  the tree-level Ward identity is 90\% of the task of satisfying the one-loop Ward identity with massive tree-level propagators; for the one-loop massless perturbative vertex, it is 100\%.  The interested reader can study \cite{corn147} for the other 10\% that arises from other complications, such the appearance of uncancelled $m^2$ in the numerator arising from the pinch process and dealing with seagull terms.

\section{\label{tweak} Tweaked $\mathbf{\phi_3^5}$ is  analogous to a d=3 NAGT}

We can get a qualitative, even semi-quantitative understanding of how the full vertex can help to tame the bulge in the propagator   by turning to some higher-dimension theories having no spin complications because they refer to scalar fields.  One of them, $\phi^3_6$, has AF analogous to that of d=4 NAGTs, and the other,   $\phi_5^3$ scalar theory in d=5, inherits certain ``wrong-sign" properties   just as does a d=3 NAGT.    These higher-dimension scalar models are to be used only at the one-dressed-loop level, and with graphical coefficients adjusted to yield appropriate results for gauge theories in two fewer dimensions. Taken seriously, the scalar  theories do not exist because there is no stable vacuum, but that will not concern us here.  Just as for NAGTs, we do not allow a bare mass term, but radiative corrections induce a mass $m$, assumed to be the same for all fields.   

There are two ways to remove the power-law divergences of the  self-energies of these theories.  One is a regulation scheme \cite{corn76} already used in d=3,4.  Appendix \ref{regsec} gives this scheme for higher dimensions.   The other way is to calculate the self-energy from a Ward identity, by introducing an Abelian charge for two of the fields, and we will concentrate on that here.
We use the corresponding Abelian vertex   to find a simple approximation to one of the scalar functions occurring in   the pole-free vertex $G_{ijk}$, as an integral over Feynman parameters.  This scalar function multiplies the Born kinematics, and its prefactor is taken not as prescribed by the d=5 model, but from the requirement that the Ward identity   yields the correct one-loop propagator in perturbation theory, given in Eq.~(\ref{1loop}).

\subsection{\label{vpapp}A vertex approximation coming from the tweaked   models  }

As shown in earlier works \cite{corn141,corn144,corn147},  the   asymptotically-free six-dimensional theory $\phi^3_6$ can be slightly modified to lead to a qualitatively-reasonable approximation for the one-dressed-loop 3-gluon vertex and gluon proper self-energy of a d=4 NAGT.  The modifications involve introducing an Abelian charge for the scalars (two of which carry equal and opposite charge) and a corresponding one-loop current vertex, as well as supplying one-loop graphs ``by hand" with coefficients taken from the NAGT.  We call this model tweaked $\phi^3_6$. The QED-like Ward identity for the tweaked-model current vertex yields the $\phi$ proper self-energy, which turns out to be practically identical to what the vertex paradigm yields for d=4 NAGTs.

Similarly, tweaked $\phi^3_5$ bears a close resemblance to a d=3 NAGT, as one might expect because the trilinear coupling $g$ has mass dimension 1/2 in both theories.  We introduce an Abelian current and find a finite propagator from its Ward identity.  In d=6, the Ward identity reduces the self-energy divergence from the quadratic divergence of $\phi^3_6$ to the logarithmic one of NAGTs.  In d=5, the Ward identity removes completely the linear self-energy divergence.  The same reduction in divergences comes from the regulator of Sec.~\ref{regsec} below.

In d=5, the one-loop Abelian current vertex is:
\begin{equation}
\label{photvert}
G_i(p_i)=(p_2-p_3)_i-2b\int\![\mathrm{d}z]\,   
 \frac{[p_2(1-2z_3)-p_3(1-2z_2)]_i}{(D+m^2)^{1/2}},
\end{equation}
with $b$ from  
the $z_j$ are Feynman parameters,   and
\begin{equation}
\label{values1}
 \int\,[dz]=2\int_0^1\,dz_1dz_2dz_3\delta(1-\sum z_i);\;D=p_1^2z_2z_3+p_2^2z_3z_1+p_3^2z_1z_2.
\end{equation}
The factor $2b$ (but not the minus sign, which comes from d=6 AF) is chosen by hand so as to give the correct perturbative correction to the propagator, as determined by the QED-like Ward identity  of Eq.~(\ref{3to5wi1}).

The current vertex should obey the Ward identity
\begin{equation}
\label{abelward}
p_{1i}G(p_i)=\Delta^{-1}(p_3)-\Delta^{-1}(p_2). 
\end{equation}
Since we are given the current vertex, this equation can be used to define the inverse propagators, provided that it has the correct structural form to be
 the difference of two inverse propagators, one with momentum $p_3$ and the other with $p_2$.  This is the case because:
\begin{equation}
\label{qedward}
p_1\cdot [p_2(1-2z_3)-p_3(1-2z_2)]=[\frac{\partial}{\partial z_2}-\frac{\partial}{\partial z_3}]
[D+m^2]
\end{equation}
and the integrals over the $z_i$ give only end-point contributions that are of the needed functional form as in Eq.~(\ref{abelward}).

In order to bridge from this Abelian one-gluon vertex to the needed 3-gluon NAGT vertex, we 
define the scalar function $G$ in the d=3 NAGT by:
\begin{equation}
\label{defg}
G_{ijk}(p_1,p_2,p_3)=\Gamma^0_{ijk}(p_1,p_2,p_3)G(p_1,p_2,p_3)
\end{equation}
where $\Gamma^0$ is the Born vertex and $\sum p_i=0$.  Similarly, we define a scalar form factor from the Abelian $\phi^3_5$ model as the coefficient of $(p_2-p_3)_i$:
\begin{equation}
\label{5to3vertex1}
G(p_1,p_2,p_3)=1-2bg^2\int\,\frac{[dz]}{(D+m^2)^{1/2}}
\end{equation}
and as the notation suggests we use the Abelian scalar form factor of this equation as an approximation to the non-Abelian scalar form factor.

Now check that the coefficient $2b$ in (\ref{photvert}) is correctly chosen.  At zero mass, the Ward identity yields:
\begin{eqnarray}
\label{3to5wi1}
\Delta^{-1}(p) & = & p^2-2bg^2\int_0^1\,dz\,(1-z)[p^2z(1-z)]^{1/2}\\ \nonumber
& = & p^2-\pi bg^2p
\end{eqnarray}
as one-loop PT perturbation theory requires (see Eq.~(\ref{1loop})).  Another feature of the approximation is that the vertex integrand $(D+m^2)^{-1/2}$
is correct for the equal-mass triangle graph in d=3.  The numerator, however, is not correct for d=3 NAGTs.  

A potential problem with the expression (\ref{5to3vertex1}) is that it might have a zero in the Euclidean region, if the mass is small enough.  Generally, such a zero leads to a tachyonic pole in the inverse propagator, from the Ward identity (\ref{gluewi}), and there is no evidence for this pole in lattice data.  Such a coincident pole and vertex zero does not appear in the S-matrix.  One way to remove the tachyon is to use a dispersion relation for a squared running charge, quite analogous to a dispersion relation used for a squared running charge proposed earlier \cite{corn141,corn144,corn147} in d=4.  This necessarily positive quantity is related, in d=4, to the vertex form factor $G(p,-p,0)$ introduced above, which cannot have any tachyon.  Of course, any running charge   in d=3 is not related to a beta-function or renormalization group; the point is that it can be defined from a plausible interpretation of the Ward identity.

\section{\label{d3runch}  A running charge in three dimensions}

In d=4 there is another use for this Abelian vertex that, at least qualitatively, links it to NAGTs.  It was argued \cite{corn144,corn147} that the current-vertex form factor that multiplies the Born kinematics yields   a running charge defined at all momenta down to zero that agrees in the UV with the usual running charge based on the renormalization group for one and two loops.    This running charge has the usual UV properties, and due to mass generation it is well-defined in the IR as well.  It is a physically well-motivated function in the IR, but   not unique.  In d=4 the renormalization group properties of PT Green's functions  suggest writing the PT propagator, multiplied by $g^2$,
\begin{equation}
\label{hruneqn}
g^2\widehat{\Delta}(p)=\bar{g}^2(p)H(p)
\end{equation}
Provided that both the PT propagator and $g^2$ are renormalized at the same renormalization point, their product is   renormalization-group-invariant \cite{corn144,corn147}.   By definition each of the factors in (\ref{hruneqn}) are also renormalization-group invariant.

Although $\widehat{\Delta}$ is unique, its factorization is not.  To be definite, in both d=3 and d=4 case we {\em define}  $H(p)$ as a standard massive propagator with a running mass:
\begin{equation}
\label{hdefinition}
H(p)= \frac{1}{p^2+m^2(p)}
\end{equation}
where the running mass is finite at zero momentum, and vanishes in perturbation theory.  In d=4 perturbation theory this definition for
  $\bar{g}$ in the UV is the usual running charge to two-loop order.  

All this has its analogs in d=3, except for the fact that there is no renormalization group in this dimension.  In d=3 we write the pole-free propagator $\Delta$ as
\begin{equation}
\label{3hruneqn}
\Delta^{-1}_{ij}(p)=\frac{p^2}{\bar{g}^2(p)}P_{ij}(p)+\dots
\end{equation}
(omitted terms are irrelevant). Then simple manipulations  of the pole-free Ward identity (\ref{gluewi}) with one vertex momentum set to zero express the so-defined running charge in the form of Eq.~(\ref{runchnum}), repeated here for convenience.
\begin{equation}
\label{runchnum2}
\bar{g}^{2}(p)=\frac{g^2}{G(p,-p,0)}
\end{equation}
where $G$ is the scalar coefficient of the Born term in the 3-gluon vertex.
Our introduction of the d=3 running charge is just another way of speaking of this vertex form factor, but we use it heuristically to argue that the vertex $G(p,-p,0)$ is everywhere positive in the Euclidean region.
 
\section{\label{regsec}  Regulating the tweaked models}

  If $\phi^3_6$ is a decent model of d=4 NAGT, does $\phi^3_5$ resemble  d=3 NAGT, with the characteristic signs inherited from d=4?  At first sight this seems impossible, since $\phi^3_5$ is not obviously a superrenormalizable theory; it has linear UV divergences in the proper self-energy as calculated directly from a Feynman graph.  But because the 3-point Abelian current vertex is finite  the Ward identity yields a finite proper self-energy.  The regulator given here   also yields a finite self-energy.

In addition, one might   worry that seagull graphs in d=3,5 are divergent.  But our   regulator \cite{corn76,corn114}   gets rid of power-law  divergences, as of course dimensional regularization also does, but the alternative regulator  does not introduce (as dimensional regularization does) potential IR divergences by the introduction of terms with massless poles.   This regulator was originally used in d=3,4 but it can be constructed for any dimension d$\geq$3.  The regulator rule is:
\begin{equation}
\label{regrule}
\int\,d^dk\,F(k^2)\rightarrow -\frac{2}{d-2}\int\,d^dk\,(1+k^2\frac{\partial}{\partial k^2})F(k^2).
\end{equation}
This replacement is an identity when the integrals in question converge, and gets rid of power-law divergences when they do not.  The effect of the regulator is to reduce divergent integrals by two space-time dimensions.  It is an alternative to using the Abelian current vertex to define the proper self-energy, but we do not pursue that subject further here.

\end{document}